\documentclass[aps,prb,twocolumn,showpacs,preprintnumbers,amsmath,amssymb]{revtex4}

\input{epsf.sty}
\usepackage{dcolumn}
\usepackage{bm}

\begin{document}

\preprint{version, currently submit to Physical Review B}

\title{Ferromagnetism and possible heavy fermion behavior in single crystals of NdOs$_4$Sb$_{12}$}

\author{P.-C. Ho, W. M. Yuhasz, N. P. Butch, N. A. Frederick, T. A. Sayles,
        J. R. Jeffries, and M. B. Maple}
\affiliation{Department of Physics and Institute for Pure and
             Applied Physical Sciences,
             University of California, San Diego,
             La Jolla, CA 92093-0360, U.S.A.}
\author{J. B. Betts and A. H. Lacerda}
\affiliation{National High Magnetic Field Laboratory/LANL,
             Los Alamos, NM 87545}

\author{P. Rogl}
\affiliation{Institut fuer Physikalische Chemie, Universitaet Wien,
             A-1090 Wien, W$\ddot{a}$hringerstr. 42, Austria}

\author{G. Giester}
\affiliation{Institut fuer Mineralogie und Kristallographie, Universitaet Wien,
             A-1090 Wien, Althanstr. 14, Austria}
\date{July 02, 2005}

\begin{abstract}
Single crystals of the filled-skutterudite compound
NdOs$_4$Sb$_{12}$ have been investigated by means of electrical
resistivity, magnetization, and specific heat measurements. The
NdOs$_4$Sb$_{12}$ crystals have the LaFe$_4$P$_{12}$-type cubic
structure with a lattice parameter of 9.3 \AA. Possible
heavy-fermion behavior is inferred from specific heat
measurements, which reveal a large electronic specific heat
coefficient \mbox{$\gamma \approx 520$\,mJ/mol-K$^2$},
corresponding to an effective mass \mbox{$m^* \approx$ 98 $m_e$}.
Features related to a ferromagnetic transition at \mbox{$\sim$
0.9\,K} can be observed in electrical resistivity, magnetization
and specific heat. Conventional Arrott-plot analysis indicates
that NdOs$_4$Sb$_{12}$ conforms to mean-field ferromagnetism.
\end{abstract}


\maketitle

\section{introduction}
The filled skutterudite compounds have the chemical formula
MT$_4$X$_{12}$, where M is an alkali (Na or K), alkaline earth
(Ca, Sr, Ba), lanthanide or actinide atom; T is a transition-metal
atom (Fe, Ru, Os); and X is a pnictogen atom (P, As, Sb). The
compounds crystallize in a LaFe$_4$P$_{12}$-type structure
with the space group Im$\bar{3}$.~\cite{Jeitschko1977} Due to the
strong hybridization between f- and conduction electrons and their
unique crystal structure, the lanthanide- and actinide-based
filled skutterudite materials display a wide range of
strongly-correlated-electron phenomena, such as BCS-like
superconductivity (e.g.,
PrRu$_4$Sb$_{12}$),~\cite{Takeda2000,Abe2002} heavy fermion
behavior (e.g., PrFe$_4$P$_{12}$),~\cite{Aoki2002} heavy fermion
superconductivity (e.g.,
PrOs$_4$Sb$_{12}$),~\cite{Maple2001,BauerED2002} ferromagnetism
(e.g., PrFe$_4$Sb$_{12}$),~\cite{BauerE2002A} metal-insulator
transitions (e.g., PrRu$_4$P$_{12}$),~\cite{Sekine1997}
Kondo-insulator behavior (e.g., UFe$_4$P$_{12}$ and
CeFe$_4$P$_{12}$),~\cite{Meisner1985} valence fluctuation behavior
(e.g., YbFe$_4$Sb$_{12}$),~\cite{Maple1999,LeitheJasper1999} and
non-Fermi-liquid behavior (e.g.,
CeRu$_4$Sb$_{12}$).~\cite{Takeda1999,BauerED2001}

Previous studies have shown ferromagnetic ordering in Nd-based
filled skutterudites. Measurements of NdFe$_4$Sb$_{12}$ revealed
ferromagnetic order below 16.5\,K with a Nd ordered moment of
2.04\,$\mu_{\rm{B}}$ and a Fe collinear moment of
0.27\,$\mu_{\rm{B}}$.~\cite{BauerE2002B} Lower ferromagnetic
transition temperatures were found in NdFe$_4$P$_{12}$ at
2\,K,~\cite{Torikachvili1987} NdRu$_4$P$_{12}$ at
1.6\,K,~\cite{Sato2003} NdRu$_4$Sb$_{12}$ at
1.3\,K,~\cite{Takeda2000,Abe2002} and NdOs$_4$Sb$_{12}$ at
0.8\,K.~\cite{Sato2003}  Since NdOs$_4$Sb$_{12}$ displays
ferromagnetism with a low Curie temperature and its neighboring
compound PrOs$_4$Sb$_{12}$ shows heavy fermion behavior and
unconventional
superconductivity,~\cite{BauerED2002,Frederick2004,Cichorek2004,Chia2003,Chia2004}
possibly involving triplet spin pairing of electrons, there is a
strong likelihood that PrOs$_4$Sb$_{12}$ is near a ferromagnetic
quantum critical point. Thus, by thoroughly characterizing the
physical properties of NdOs$_4$Sb$_{12}$, a deeper insight into
the unusual behavior of PrOs$_4$Sb$_{12}$ may be attained. Earlier
studies of the compound NdOs$_4$Sb$_{12}$ only reported the
results of structural refinement~\cite{Evers1995} and the value of
the Curie temperature.~\cite{Sato2003}  In this report, we present
a new and detailed investigation of NdOs$_4$Sb$_{12}$ single
crystals, including measurements of X-ray diffraction, electrical
resistivity, magnetization, and specific heat.  We also discuss
possible heavy fermion behavior in this Nd-based filled
skutterudite compound.

\section{Experimental Details}
NdOs$_4$Sb$_{12}$ single crystals were grown in a molten Sb flux
as described previously,~\cite{Frederick2004} using high purity Nd,
Os ($3.5$N), and Sb ($6$N).  X-ray powder diffraction
measurements were performed with a Rigaku D/MAX B x-ray machine on
a powder prepared by grinding several single crystals, which
indicated single phase NdOs$_4$Sb$_{12}$ with a minor impurity
peak of Sb \mbox{$(\lesssim 10 \%)$}. The crystals had a
LaFe$_4$P$_{12}$-type BCC structure,~\cite{Jeitschko1977} with a
lattice parameter $a = 9.30$\,\AA. Two single crystals of similar
dimension were selected for single crystal X-ray diffraction measurements.
The data were collected on a four-circle Nonius Kappa diffractometer
at 296\,K using Mo K$_{\alpha}$ radiation ($\lambda$ = 0.071073\,nm). 
No absorption corrections were necessary because of the rather
regular crystal shape and small dimensions of the investigated
crystals. The structure was refined with the SHELXS-97
program.~\cite{SHELXS97}

Electrical resistivity $\rho(T,H)$ was measured using the standard
4-wire technique in a Quantum Design PPMS system and in a
$^3$He-$^4$He dilution refrigerator in fields up to 8\,T. The low
temperature (0.02\,K - 2.6\,K) and high-field (8\,T - 18\,T)
$\rho(T,H)$ measurements were performed in the National High
Magnetic Field Laboratory at Los Alamos National Laboratory. The
electrical current applied to the sample was perpendicular to the
applied magnetic field, which was along the [001] direction, in
all $\rho(T,H)$ measurements. Measurements of $\rho(T,P)$ were
made under hydrostatic pressures up to 28\,kbar in a
beryllium-copper piston-cylinder clamp~\cite{BauerED2001A} and a
$^{4}$He cryostat. The pressure was determined inductively from
the pressure-dependent superconducting transition of a Pb
manometer.

DC magnetic susceptibility from 2\,K to 300\,K was measured in a
Quantum Design MPMS SQUID magnetometer. The magnetization $M(H,T)$
measurements were carried out in a $^3$He Faraday magnetometer
with a gradient field of  $\sim$ 0.05 - 0.1\,T/cm in external
fields up to 5.5\,T and at temperatures between 0.4\,K and 2\,K.
For the Faraday magnetometer measurements, several single crystals
(total mass of 21.3\,mg) were combined in a mosaic fashion and
measurements were performed with magnetic field applied along the
[001] axis. Specific heat $C(T)$ of multiple single crystals
(total mass of 42.15\,mg) was measured between 0.5\,K and 70\,K in
a $^3$He calorimeter using a semi-adiabatic heat-pulse technique.

\section{Results and Discussion}

\subsection{Single crystal structural refinement}
\squeezetable
\begin{table*}
\caption{Single crystal structural data measured at $T = 296$\,K
         for NdOs$_4$Sb$_{12}$.  The crystal structure is LaFe$_4$P$_{12}$-type
         with the space group Im$\bar{3}$ (No. 204).
         The range of X-ray scattering angle is
         \mbox{2$^{\circ}$ $< 2 \theta$ $<$ 80$^{\circ}$}.}
\begin{tabular}{|lc|lc|lc|}

 \hline
 \multicolumn{6} {|c|}{\bf NdOs$_4$Sb$_{12}$} \\  \hline
 Crystal size & 64 $\times$ 78 $\times$ 84 $\mu$m$^3$
  & Lattice parameter $a$ [$\rm{\AA}$] & 9.3075(2)
  & Density $\rho$ [g/cm$^3$] & 9.745 \\\hline
 Reflection in refinements & 473 $\leq$ 4 $\sigma$(F$_0$) of 482
  & Number of variables & 11
  & R$_F^2 = \sum \vert F_0^2 - F_c^2 \vert / \sum F_0^2$ & 0.0186 \\\hline
 Goodness of fit &  1.255
  &  &
  &  & \\\hline
 Nd in 2a (0, 0, 0); &
  & Thermal displacement & [$\rm{\AA}^2$]
  & Interatomic distances [$\rm{\AA}$] & \\
 Occupancy & 1.00(1)
  & Nd: $U_{11}$ = U$_{22}$ = U$_{33}$ & 0.0482(5)
  & Nd - 12 Sb & 3.4831 \\\hline
 Os in 8c (1/4, 1/4, 1/4); &
  & Thermal displacement & [$\rm{\AA}^2$]
  & Interatomic distances [$\rm{\AA}$] & \\
 Occupancy & 1.00(1)
  & Os: $U_{11}$ = U$_{22}$ = U$_{33}$ & 0.0025(1)
  & Os - 6 Sb & 2.6239 \\\hline
 Sb in 24g (0, y, z);~~y: & 0.15597(3)
  & Thermal displacement & [$\rm{\AA}^2$]
  & Interatomic distances [$\rm{\AA}$] &  \\
  ~~~~~~~~~~~~~~~~~~~~~~~~~~z: & 0.34017(3)
  & Sb: $U_{11}$ & 0.0026(1)
  & Sb - 1 Sb & 2.9033 \\
  Occupancy & 1.00(1)
  &~~~~~\,$U_{22}$ & 0.0044(1)
  & ~~~~\,- 1 Sb & 2.9752 \\
 &
  & ~~~~~\,$U_{33}$ & 0.0065(1)
  & ~~~~\,- 2 Os & 2.6239 \\
 &
  &  &
  & ~~~~\,- 1 Nd & 3.4831 \\\hline

\end{tabular}
\label{tbl:RoglXrayresult}
\end{table*}

Structural refinement was performed on X-ray diffraction data
collected from single crystals of NdOs$_4$Sb$_{12}$; the results
are listed in Table~\ref{tbl:RoglXrayresult}. The thermal
displacement parameters $U_{ii}$ of the Nd atoms are isotropic and
have large values compared with the $U_{ii}$ for the Os and Sb
atoms, a common feature in the filled skutterudites.  The Nd sites
are fully occupied, which is not always the case for filled
skutterudites, such as Pr$_{0.73}$Fe$_4$Sb$_{12}$ and
Eu$_{0.95}$Fe$_4$Sb$_{12}$.~\cite{BauerE2002A,BauerED2004A} If the
NdOs$_4$Sb$_{12}$ crystal is considered to be a simple Debye solid
with the Nd atoms behaving like Einstein oscillators, the thermal
displacement and the Einstein temperature $\Theta_{\rm{E}}$ are
related by
\begin{equation}
 U = \frac{\hbar^2}{2m_{_{\rm{Nd}}}k_{\rm{B}}\Theta_E}
     coth \left( \frac{\Theta_{\rm{E}}}{2T} \right),
\end{equation}
where $m_{_{\rm{Nd}}}$ is the atomic mass of Nd.  For
NdOs$_4$Sb$_{12}$, $\Theta_{\rm{E}}$ is estimated as \mbox{$\sim$
45\,K}, which is close to the values found for thallium-filled
antimony skutterudites such as Tl$_{0.22}$Co$_{4}$Sb$_{12}$,
Tl$_{0.5}$Co$_{3.5}$Fe$_{0.5}$Sb$_{12}$,
Tl$_{0.8}$Co$_{3}$FeSb$_{12}$, and
Tl$_{0.8}$Co$_{4}$Sb$_{11}$Sn.~\cite{Sales1999,Hermann2003}

\subsection{Magnetic Properties}
\epsfxsize=230pt \epsfysize=230pt
\begin{figure}
 \epsfbox{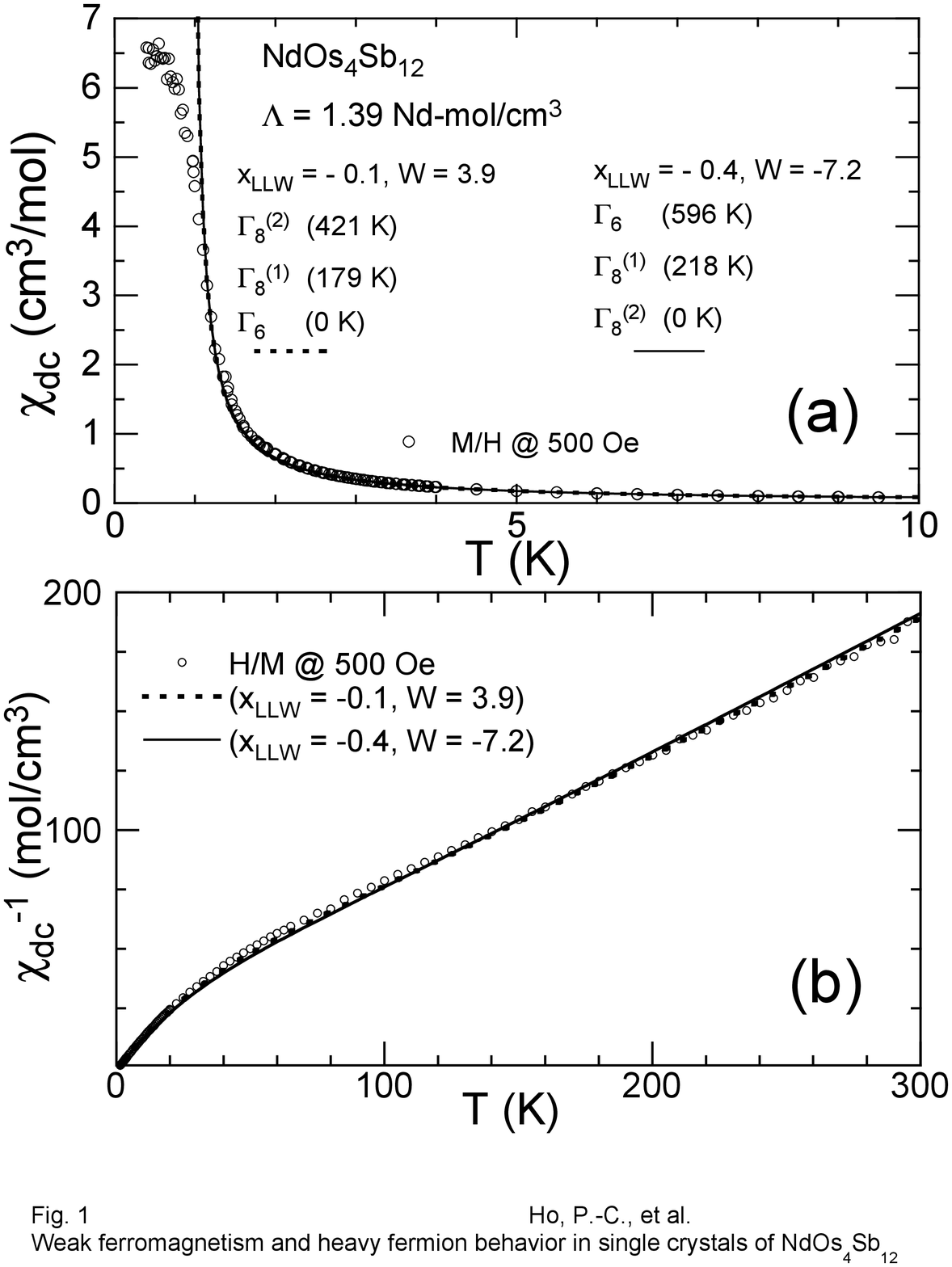}
 \caption{(a) DC magnetic susceptibility $\chi_{\rm{dc}}$ vs temperature $T$ from 0\,K to 10\,K
              measured at 500\,Oe.  Fits of $\chi_{\rm{dc}}(T)$ to a CEF model in which the
              ground state is either $\Gamma_6$ (dashed line) or $\Gamma_8^{(2)}$ (solid line).
              Below 10\,K, these two lines overlap.
          (b) $\chi_{\rm{dc}}^{-1}$ vs $T$ from 0\,K to 300\,K.
              Fits of $\chi_{\rm{dc}}(T)$ and $\chi_{\rm{dc}}^{-1}(T)$ to a CEF model in which
              the ground state is either $\Gamma_6$ (dashed line) or $\Gamma_8^{(2)}$ (solid
              line).}
 \label{fig:chivsTwCEFfit}
\end{figure}

\begin{figure}
 \epsfbox{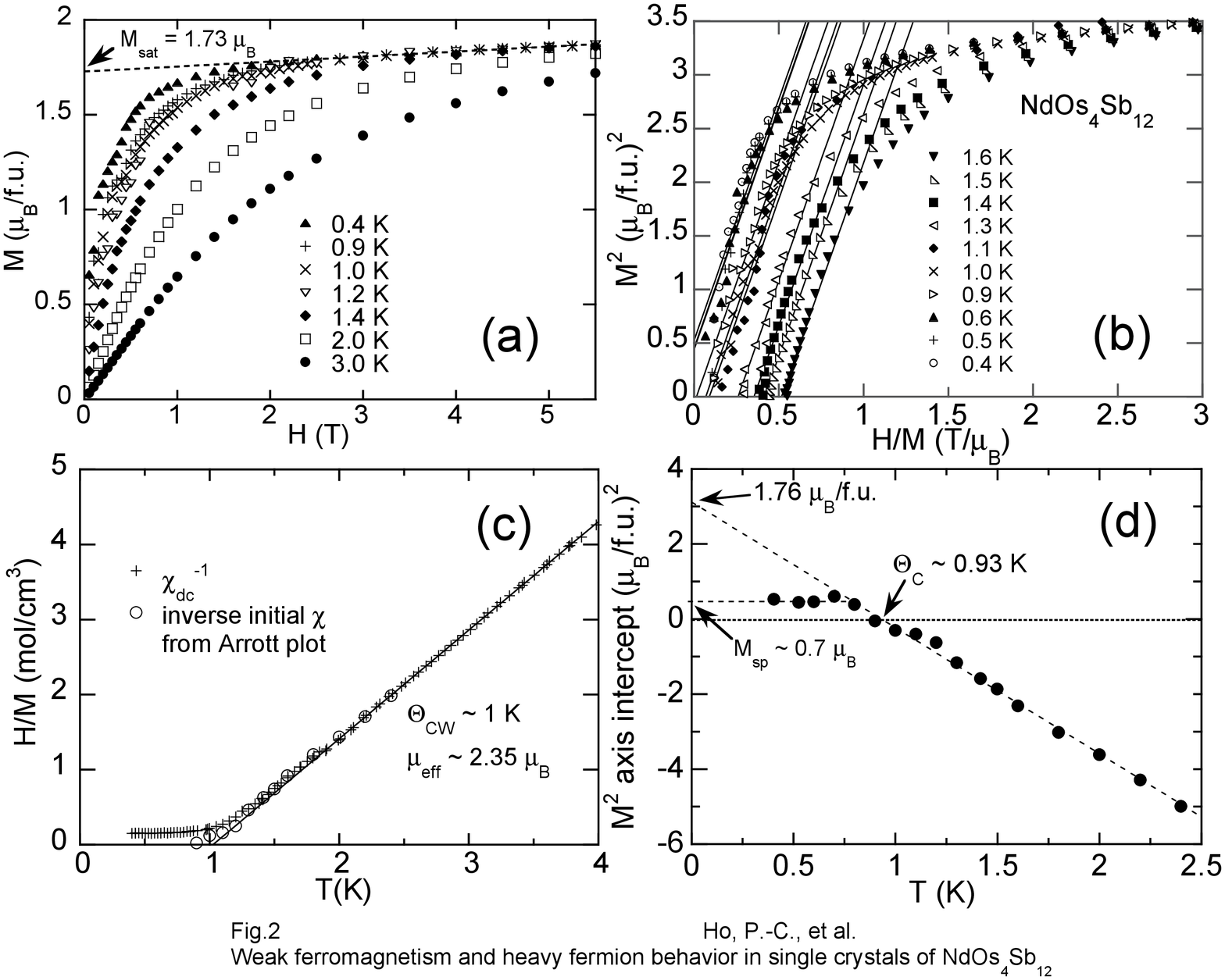}
 \caption{(a) Magnetization $M$ vs magnetic field $H$ isotherms and the saturation
              magnetization $M_{\rm{sat}}$ determined
              from $M$ vs $H$ isotherms below the Curie temperature $\Theta_{\rm{C}}$.
          (b) $M^2$ vs $H/M$ isotherms (Arrott plot) for NdOs$_4$Sb$_{12}$.
          (c) Inverse initial magnetic susceptibility (open circles)
              determined from \mbox{$M^2 = 0$} intercepts of  $M^2$ vs $H/M$ isotherms,
              compared to the $\chi_{\rm{dc}}^{-1}(T)$ data (pluses).
              The line is a Curie-Weiss fit.
          (d) \mbox{$H/M = 0$} intercepts of $M^2$ vs $H/M$ isotherms plotted versus $T$.
              Positive values correspond to the spontaneous magnetization $M_{\rm{sp}}$.}
 \label{fig:ArrottPlot}
\end{figure}

The dc magnetic susceptibility $\chi_{\rm{dc}}(T)$ of
NdOs$_4$Sb$_{12}$ was measured at 500\,Oe and is displayed in
Fig.~\ref{fig:chivsTwCEFfit}(a). The $\chi_{\rm{dc}}^{-1}(T)$ data
(Fig.~\ref{fig:chivsTwCEFfit}(b)) exhibit different slopes at high
and low temperatures. The linear slope of $\chi_{\rm{dc}}^{-1}(T)$
between 65\,K and 300\,K yields a Curie constant (C$_{\rm{CW}}$
$\equiv$ \mbox{(N$_{\rm{A}}$$\mu_{\rm{eff}}^2$)/(3$k_{\rm{B}})$})
$\sim 1.85$\,cm$^3$K/mol, a negative Curie-Wiess temperature
($\Theta_{\rm{CW}}$) $\sim$ -43\,K, and a effective moment
$\mu_{\rm{eff}} \sim$ \mbox{3.84\,$\mu_{\rm{B}}$}, which is close
to the Nd$^{3+}$ free-ion value of \mbox{3.62\,$\mu_{\rm{B}}$}.
The Curie-Weiss fit to the low-temperature range of
$\chi_{\rm{dc}}^{-1}(T)$ (Fig.~\ref{fig:ArrottPlot}(c)) gives
C$_{\rm{CW}}$ $\sim 0.69$\,cm$^3$K/mol, a positive
$\Theta_{\rm{CW}}$ $\sim$ 1\,K, and a value of $\mu_{\rm{eff}}
\sim$ \mbox{2.35\,$\mu_{\rm{B}}$}. The curvature in
$\chi_{\rm{dc}}^{-1}(T)$ is due to the influence of the
crystalline electric field (CEF), and the positive
$\Theta_{\rm{CW}}$ from the low-temperature fit indicates
ferromagnetic order developing below 1\,K.

Although the crystal structure of NdOs$_4$Sb$_{12}$ has
tetrahedral symmetry (T$_h$),~\cite{Takegahara2001} this is only a
slight deviation from cubic symmetry (O$_h$). Thus, to simplify
the CEF analysis, only O$_h$ symmetry was considered. In an ionic
(localized) model with cubic symmetry, the Nd$^{3+}$ ten-fold
degenerate \mbox{J = 9/2} Hund's rule ground state multiplet
splits into a $\Gamma_6$ doublet and two $\Gamma_8$
($\Gamma_8^{(1)}, \Gamma_8^{(2)}$) quartet states. In the
treatment of Lea, Leask, and Wolf,~\cite{LLW1962} these energy
levels and their corresponding wave functions in cubic O$_h$
symmetry can be parameterized by the variables $x_{\rm{LLW}}$ and
$W$, where $x_{\rm{LLW}}$ is the ratio of the fourth- and
sixth-order terms of the angular momentum operators and $W$ is an
overall energy scale. The CEF contribution to the molar magnetic
susceptibility can be determined from the
expression~\cite{Fulde1979}
\begin{equation}
\chi_{\rm{CEF}}(T) = N_{\rm{A}}g_{_{J}}^2\mu_{\rm{B}}^2 \left[
    \sum_{i}\frac{\vert\langle i|J_z|i\rangle\vert^{2}}{k_{\rm{B}}T}p_{i}
    -2\sum_{i,j(\neq i)}\frac{\vert\langle i|J_z|j\rangle\vert^{2}}{E_{i}-E_{j}}p_{i}
    \right],
\end{equation}
where $N_{\rm{A}}$ is Avogadro's number, $g_{_{J}}$ is the
Land$\acute{\rm{e}}$ g-factor, $\mu_{\rm{B}}$ is the Bohr
magneton, $p_i = e^{-E_i/(k_{\rm{B}}T)}/Z$ is the thermal
population probability ($Z$ is the partition function), and the
$E_{i}$'s are the energies of the multiplets.  However, the
occurrence of ferromagnetic order in NdOs$_4$Sb$_{12}$ requires
the presence of a molecular field constant $\Lambda$ to account
for the effective field, with $\chi^{-1} = \chi_{\rm{CEF}}^{-1} -
\Lambda$.

The low-temperature value of $\mu_{\rm{eff}}$ (\mbox{2.35
$\mu_{\rm{B}}$}) indicates that the ground state of Nd$^{3+}$ is
either $\Gamma_6$ or $\Gamma_8^{(2)}$. A wide range of
$x_{\rm{LLW}}$ values with a $\Gamma_6$ ground state can fit the
$\chi(T)$ data reasonably well ($-1 \leq x_{\rm{LLW}} \leq 0.1$).
For a $\Gamma_8$ ground state, a good fit to the $\chi(T)$ data is
only found for \mbox{$x_{\rm{LLW}} \approx -0.4$}. All the fits
imply that the splitting between the ground and first excited
states is greater than 120\,K. When compared with the $\rho(T)$
data (to be discussed later), the best fits are (I)
\mbox{$x_{\rm{LLW}} = -0.1$}, \mbox{$W$ = 3.9}, and (II)
\mbox{$x_{\rm{LLW}} = -0.4$}, \mbox{$W = -7.2$}, corresponding to
the following level schemes: (I) \mbox{$\Gamma_6$ (0\,K)},
\mbox{$\Gamma_8^{(1)}$ (180\,K)}, \mbox{$\Gamma_8^{(2)}$ (420\,K)}
and (II) \mbox{$\Gamma_8^{(2)}$ (0\,K)}, \mbox{$\Gamma_8^{(1)}$
(220\,K)}, \mbox{$\Gamma_6$ (600\,K)}, with \mbox{$\Lambda$ = 1.39
Nd-mol/cm$^3$} and an effective ground state moment $\sim$
2.31$\mu_{\rm{B}}$ for both schemes. These fits are displayed in
Figs.~\ref{fig:chivsTwCEFfit}(a) and (b). The Heisenberg
interaction strength between a Nd ion and its eight nearest
neighbors is estimated to be 0.522\,K from $\Lambda$.

\epsfxsize=220pt \epsfysize=225pt

Isothermal magnetization measurements $M(H)$, displayed in
Fig.~\ref{fig:ArrottPlot}(a), were made above and below the Curie
temperature ($\Theta_{\rm{C}}$).  In the paramagnetic state, the
$M(H)$ isotherms between 2\,K and 5\,K can be fit well by a
Brillouin function with effective $J = 2.7$ that includes a
temperature shift of 1\,K due to ferromagnetism.  The $M(H)$
isotherms in the vicinity of $\Theta_{\rm{C}}$ were used to
construct a conventional Arrott plot consisting of $M^2$ vs $H/M$
isotherms, shown in Fig.~\ref{fig:ArrottPlot}(b). The isotherms in
the Arrott plot are linear and parallel in the vicinity of
$\Theta_{\rm{C}}$ for H $\leq$ 1\,T, indicating that
NdOs$_4$Sb$_{12}$ is a mean-field ferromagnet. The intercepts of
the linear fits to the $(H/M)$ axis ($\equiv$ the inverse initial
magnetic susceptibility) agree well with the low-temperature
$\chi_{\rm{dc}}^{-1}(T)$ data (Fig.~\ref{fig:ArrottPlot}(c)). The
intercepts of the linear fit to the $M^2$ axis are shown in
Fig.~\ref{fig:ArrottPlot}(d), where zero identifies
$\Theta_{\rm{C}} = 0.93$\,K. Below $\Theta_{\rm{C}}$, the
intercept of the $M^2$-axis corresponds to the square of the
spontaneous magnetization ($M_{\rm{sp}}^2$), which levels off and
results in a small value of \mbox{$M_{\rm{sp}}$ $\sim 0.7
\mu_{\rm{B}}$/f.u.}. However, a linear extrapolation  of the
negative $M^2$-axis intercept back to zero temperature yields a
much larger value of \mbox{$\sim$ 1.76 $\mu_{\rm{B}}$/f.u.}, which
is comparable to the saturation magnetization $M_{\rm{sat}}$ of
\mbox{$\sim$ 1.73 $\mu_{\rm{B}}$/f.u.} determined directly from
the $M(H)$ data of NdOs$_4$Sb$_{12}$
(Fig.~\ref{fig:ArrottPlot}(a)). The value of $M_{\rm{sat}}$ is
also consistent with the saturation magnetization found in
NdFe$_4$P$_{12}$
(1.72\,$\mu_{\rm{B}}$/f.u.).~\cite{Torikachvili1987}

\subsection{Electrical Resistivity}
\epsfxsize=225pt \epsfysize=220pt
\begin{figure}
 \epsfbox{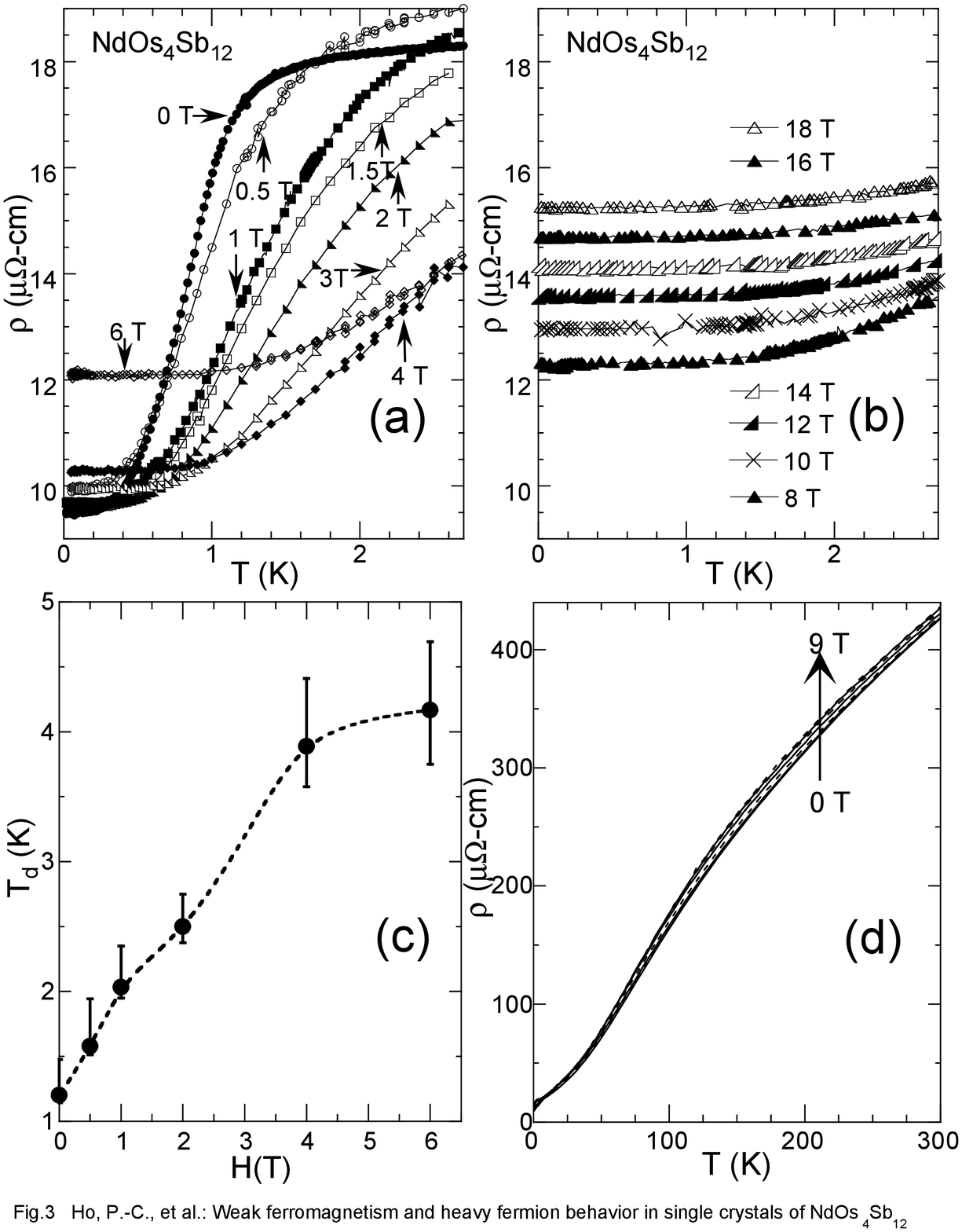}
 \caption{(a) Low-temperature electrical resistivity $\rho$ vs temperature $T$ for magnetic
              fields $H$ between 0\,T and 6\,T for NdOs$_4$Sb$_{12}$. The solid lines are
              guides to the eye.
           (b) Low-temperature electrical resistivity $\rho$ vs $T$ for magnetic fields
              $H$ between 8\,T and 18\,T for NdOs$_4$Sb$_{12}$. The solid lines are guides
              to the eye.
          (c) Position of the shoulder $T_{\rm{d}}$ in $\rho(T)$ (which corresponds to
              $\Theta_{\rm{C}}$) vs $H$, with vertical bars indicating the width of the
              transition at $T_{\rm{d}}$ as described in the text.
          (d) High-$T$ resistivity $\rho$ vs $T$ at $H$ = 0\,T, 0.5\,T, 1\,T, 2\,T,
              4\,T, 6\,T, 8\,T, 9\,T.}
 \label{fig:rhovsTinH}
\end{figure}

Low-temperature electrical resistivity $\rho(T)$ data for
NdOs$_4$Sb$_{12}$ in various magnetic fields from 0\,T to 18\,T
are shown in Figs.~\ref{fig:rhovsTinH}(a) and (b). The zero-field
residual resistivity ratio \mbox{RRR $\equiv
\rho(300\,$K)$/\rho(0.02$K)} of $\sim$ 45 indicates that the
single crystal studied is of good metallurgical quality
(Fig.~\ref{fig:rhovsTinH}(a) and (d)). A shoulder occurs in the
zero-field $\rho(T)$ curve at $\sim$ 1.2\,K, below which $\rho(T)$
has a sharp drop, indicating the development of an ordered state.
The temperature $T_{\rm{d}}$ at which this drop occurs is defined
as the intercept of two lines, one of which is a linear fit to the
data above the transition while the other is a linear fit to the
data below the transition. The upper and lower limits of the
transition are defined as the temperatures midway between
$T_{\rm{d}}$ and the temperatures at which the data deviate from
the linear fits. Shown in Fig.~\ref{fig:rhovsTinH}(c) is the field
dependence of $T_{\rm{d}}$, which increases with increasing field
up to 6\,T, and is no longer observed in the $\rho(T)$ data at
higher fields. The temperature $T_{\rm{d}}$ correlates with the
onset of ferromagnetic order at 0.9\,K determined from
magnetization and specific heat measurements. Displayed in
Fig.~\ref{fig:rhovsTinH}(d) are the high-temperature
\mbox{$\rho(T$, 0\,T $\leq H \leq$ 9\,T)} data for
NdOs$_4$Sb$_{12}$, which show a slight increase in $\rho(T)$ with
increasing field.

\epsfxsize=250pt \epsfysize=198pt
\begin{figure}
 \epsfbox{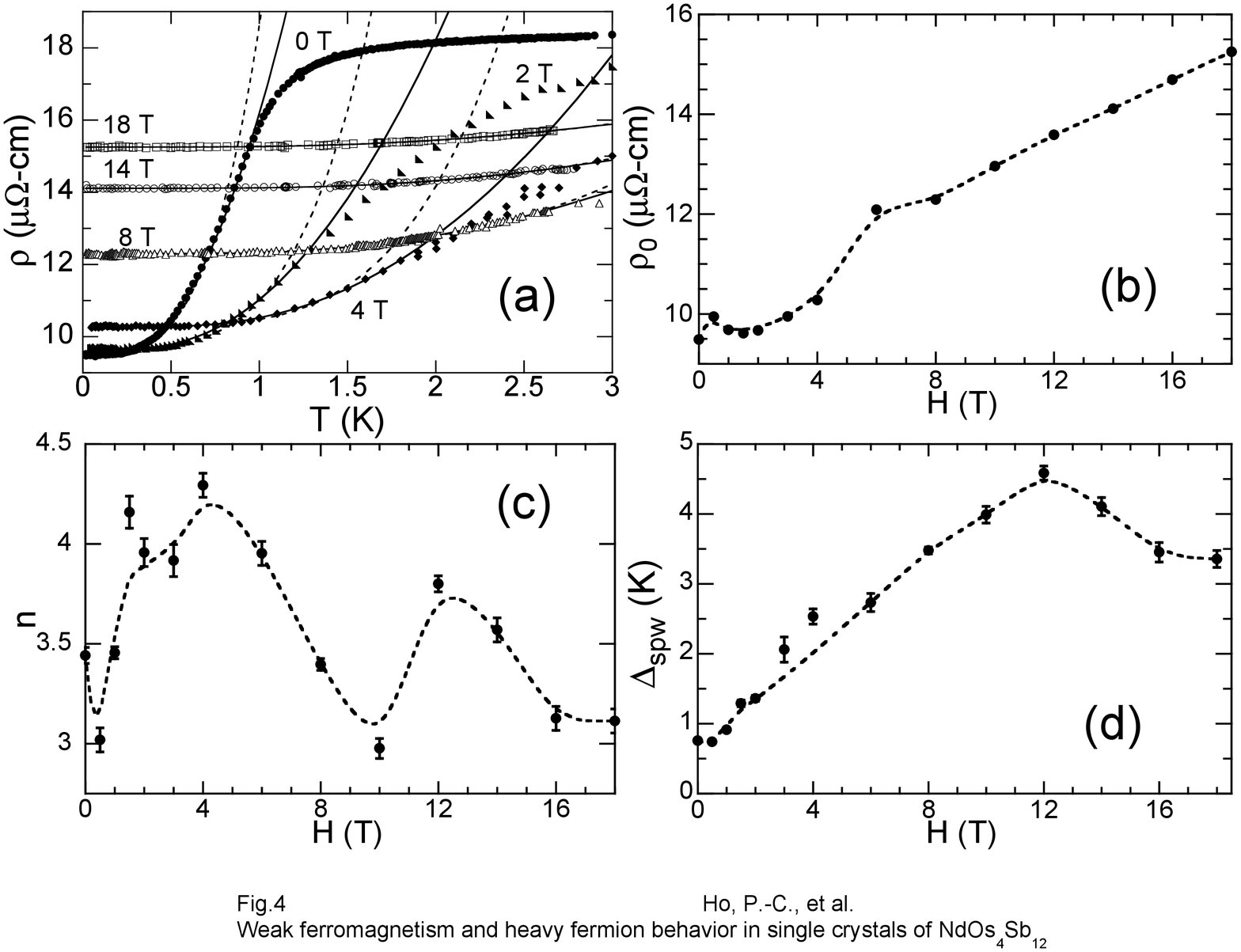}
 \caption{(a) Low-$T$ electrical resistivity $\rho$ vs temperature $T$ with power-law
              (dashed line) and spin-wave (solid line) fits at various magnetic fields
              for NdOs$_4$Sb$_{12}$.
          (b) Residual resistivity $\rho_0$ vs $H$.
          (c) Exponent $n$ of the power-law fit vs $H$.
          (d) Energy gap $\Delta_{\rm{spw}}$ from the spin-wave fit vs $H$.}
 \label{fig:SpwPwrFits}
\end{figure}

In order to analyze the behavior of the resistivity below
$\Theta_{\rm{C}}$, the $\rho(T)$ data were fit with a power-law of
the form $\rho(T) = \rho_0 + BT^n$. The fitting curves are plotted
as dashed lines in Fig.~\ref{fig:SpwPwrFits}(a). The residual
resistivity $\rho_0$ increases with increasing field and has a
linear $H$-dependence above 8\,T (Fig.~\ref{fig:SpwPwrFits}(b)).
Between 0\,T and 18\,T, the exponent $n$ varies from 3 to 4, which
indicates that NdOs$_4$Sb$_{12}$ exhibits neither typical
Fermi-liquid ($n$ $\sim$ 2) nor typical non-Fermi-liquid ($n$ $<$
2) behavior (Fig.~\ref{fig:SpwPwrFits}(c)). Since ferromagnetic
ordering occurs below $\Theta_{\rm{C}}$, electron-spin wave
scattering was considered with the form~\cite{Andersen1980}
\begin{equation}
\rho(T)= \rho_0 +
A\frac{T}{\Delta_{\rm{spw}}}\left(1+2\frac{T}{\Delta_{\rm{spw}}}\right)
\exp\left(-\frac{\Delta_{\rm{spw}}}{T}\right),
\end{equation}
where $\Delta_{\rm{spw}}$ is the spin wave energy gap, which may
result either from magnetic anisotropy or from broken symmetry due
to presence of a CEF. This formula describes the $\rho(T,H)$ data
well, and the fitting curves are shown as solid lines in
Fig.~\ref{fig:SpwPwrFits}(a).  As determined from these fits, the
spin-wave energy gap $\Delta_{\rm{spw}}$ is $\sim$ 0.75\,K at
0\,T, increases approximately linearly to $\sim$ 4.5\,K as the
field increases to 12\,T, and then drops to $\sim$ 3.4\,K at 18\,T
(Fig.~\ref{fig:SpwPwrFits}(d)).

\epsfxsize=250pt \epsfysize=230pt
\begin{figure}
 \epsfbox{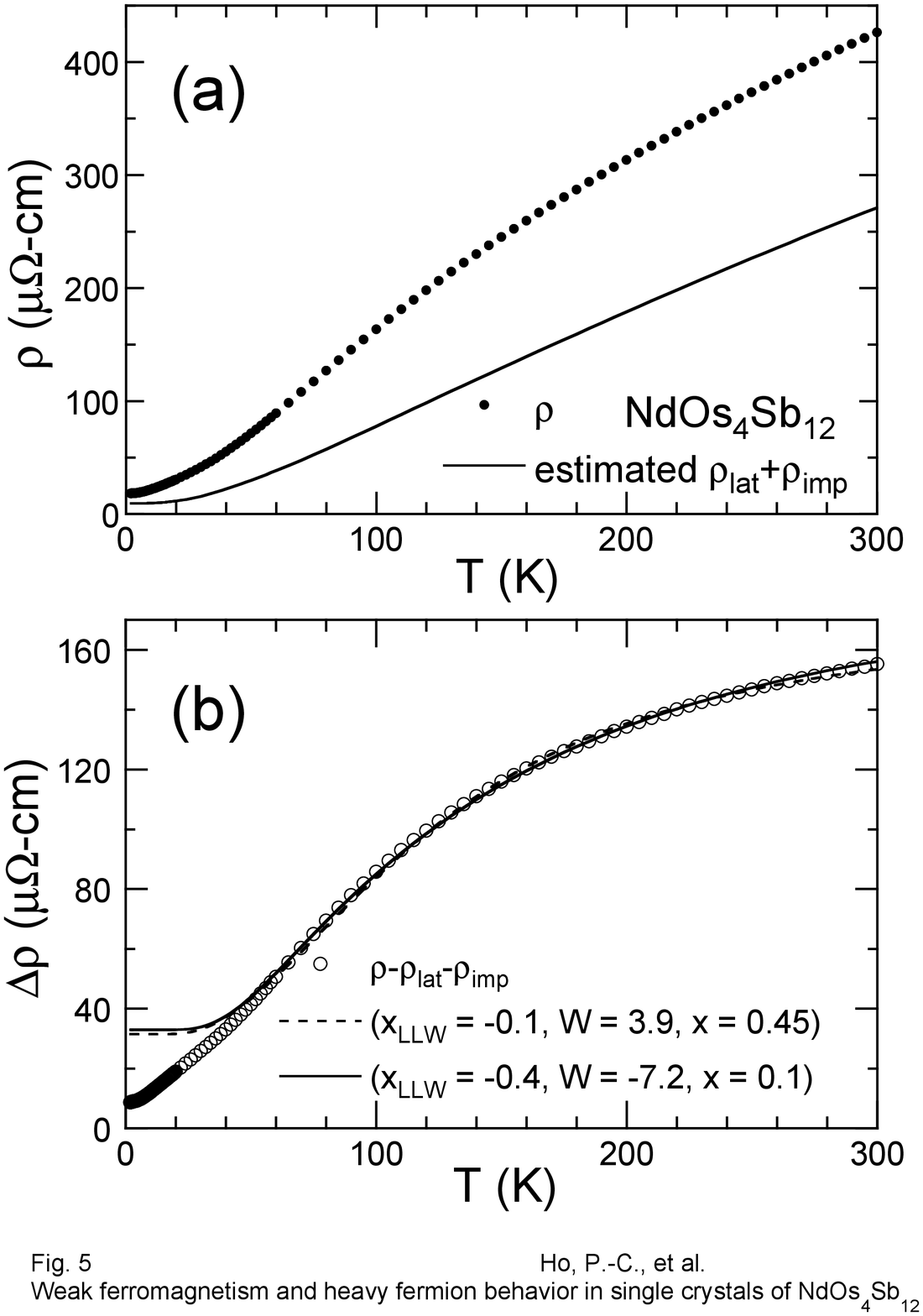}
 \caption{(a) Zero-field resistivity $\rho$
              and the estimated $\rho_{\rm{lat}} + \rho_{\rm{imp}}$ vs temperature $T$ for
              NdOs$_4$Sb$_{12}$, where \mbox{$\rho_{\rm{imp}} \sim$ 9.4 $\mu\Omega$-cm}.
          (b) Temperature dependence of the incremental resistivity
              \mbox{$\Delta\rho = \rho -\rho_{\rm{lat}} -\rho_{\rm{imp}}$} and
              CEF fits for two different ground states: $\Gamma_6$ (dashed line)
              and $\Gamma_8^{(2)}$ (solid line), where $x/(1-x)$ is the ratio of s-f exchange
              to aspherical Coulomb scattering for NdOs$_4$Sb$_{12}$.}
 \label{fig:rhoT@0TwCEFfit}
\end{figure}

Figure~\ref{fig:rhoT@0TwCEFfit}(a) displays the zero-field
$\rho(T)$ data, which have a slight negative curvature at $\sim$
130\,K that may be related to scattering from the CEF levels. In
order to analyze the CEF contribution to $\rho(T)$ at 0\,T, it is
necessary to subtract a lattice contribution $\rho_{\rm{lat}}(T)$
and an impurity contribution $\rho_{\rm{imp}}$ (\mbox{$\sim$
9.4\,$\mu\Omega$-cm}) from the resistivity data. Usually,
$\rho_{\rm{lat}}(T)$ is estimated from an isostructural
nonmagnetic reference compound; in the case of NdOs$_4$Sb$_{12}$,
the resistivity of LaOs$_4$Sb$_{12}$, which has an empty 4-f
shell, was used. However, above 100\,K, $\rho(T)$ of
LaOs$_4$Sb$_{12}$ exhibits a significant negative curvature that
is common in La-based compounds such as
LaAl$_2$.~\cite{Slebarski1985,Maple1969} This curvature is
generally less pronounced in Sc-, Y-, and Lu-based compounds,
which have completely empty or filled 4f-electron shells. However,
the compounds ScOs$_4$Sb$_{12}$, YOs$_4$Sb$_{12}$ and
LuOs$_4$Sb$_{12}$ have not yet been synthesized.  Above 100\,K,
$\rho_{\rm{lat}}(T)$ was estimated from SmOs$_4$Sb$_{12}$, as
$\rho(T)$ of SmOs$_4$Sb$_{12}$ has an approximately linear-$T$
dependence between 100\,K and 300\,K.~\cite{Yuhasz2004} The s-f
exchange scattering effect in $\rho(T)$ of SmOs$_4$Sb$_{12}$ only
appears below 80\,K due to the small energy splitting
\mbox{($\sim$ 35\,K)} between the ground and the  first excited
states in that compound. Thus it is reasonable to assume that
\mbox{$\rho(T)$} of SmOs$_4$Sb$_{12}$ for \mbox{100\,K $\le T \le$
300\,K} is due to electron-phonon scattering and use it as an
estimate of the high-temperature portion of $\rho_{\rm{lat}}(T)$
for NdOs$_4$Sb$_{12}$.

\begin{figure}
 \epsfbox{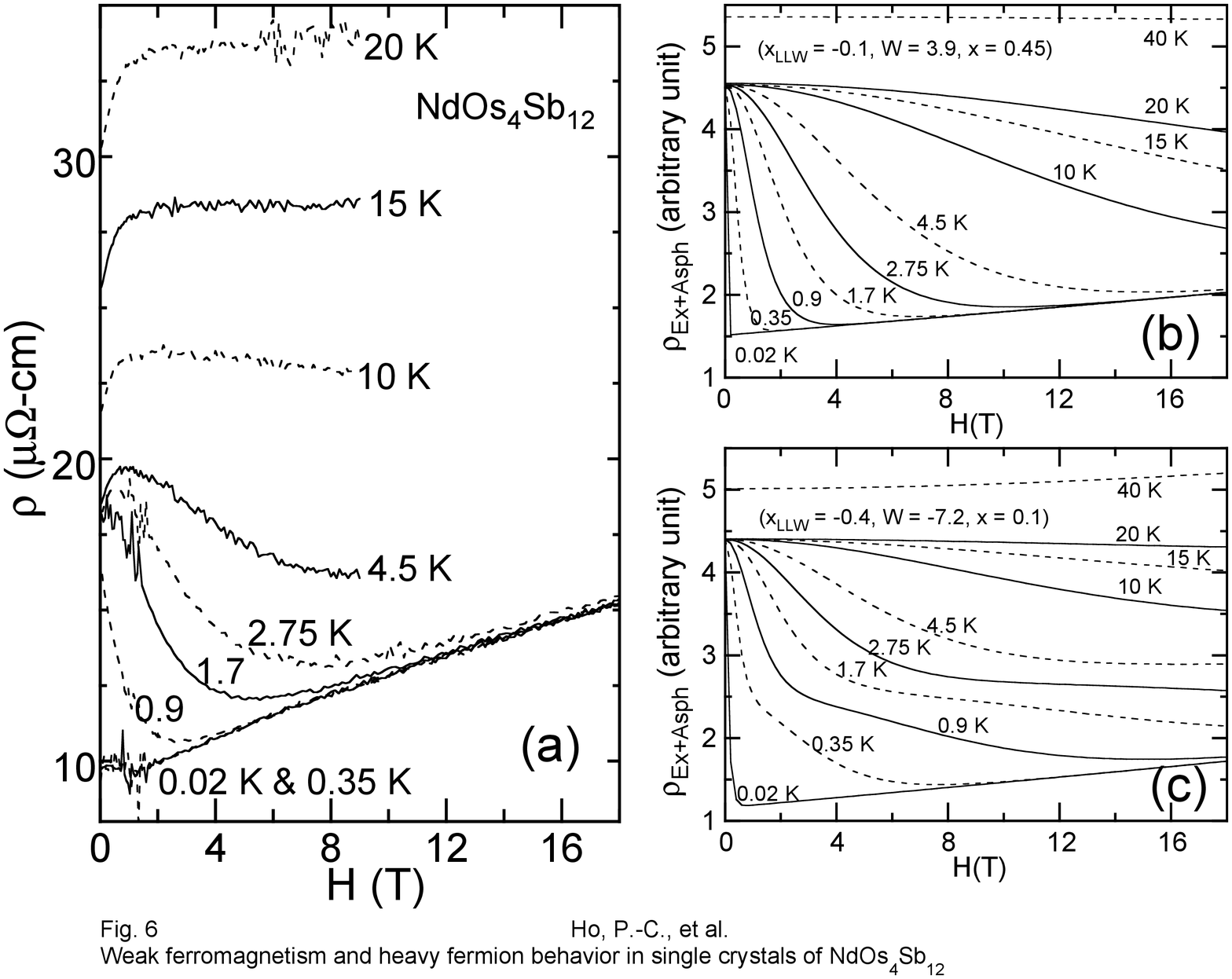}
 \caption{(a) Isotherms of electrical resistivity $\rho$ vs magnetic field $H$
              for NdOs$_4$Sb$_{12}$.
          (b) and (c) Calculated $\rho$ vs $H$ isotherms including the effects of
              s-f exchange and aspherical Coulomb scattering using the CEF parameters
              determined from zero-field data.}
 \label{fig:rhovsH}
\end{figure}

The incremental resistivity $\Delta\rho(T) = \rho(T) -
\rho_{\rm{lat}}(T) - \rho_{\rm{imp}}$
(Fig.~\ref{fig:rhoT@0TwCEFfit}(b)) is best described by two
energy-level schemes that are consistent with the CEF analysis of
$\chi(T)$ discussed previously. During the CEF analysis of
$\Delta\rho(T)$, it was also found that s-f exchange scattering
alone could not entirely account for $\Delta\rho(T)$; otherwise,
$\rho_{\rm{imp}}$ would always be negative, which is unphysical.
Thus, the effect of aspherical Coulomb
scattering~\cite{Elliott1954,Fulde1972,Frederick2003} was also
considered, with the ratio between the s-f exchange scattering and
the aspherical Coulomb scattering defined as \mbox{$x$ : (1-$x$)}.
In Fig.~\ref{fig:rhoT@0TwCEFfit}(b), the fit curves representing
the two best-fit energy schemes are plotted in comparison with
$\Delta\rho(T)$, which is described well by both fits above 40\,K.
The departure of $\Delta\rho(T)$ from the fits below 40\,K may be
due to a reduction of the electron scattering, resulting from the
development of the coherent heavy-fermion ground state in
NdOs$_4$Sb$_{12}$ as the temperature is lowered. The $\rho(H)$
isotherms are displayed in Fig.~\ref{fig:rhovsH}(a) and
qualitatively agree with the $\rho(H)$ isotherms generated using
the cubic-CEF parameters determined from the zero-field fits to
$\Delta\rho(T)$ (Figs.~\ref{fig:rhovsH}(b) and (c)).

\epsfxsize=250pt \epsfysize=200pt

\epsfxsize=240pt \epsfysize=240pt
\begin{figure}
 \epsfbox{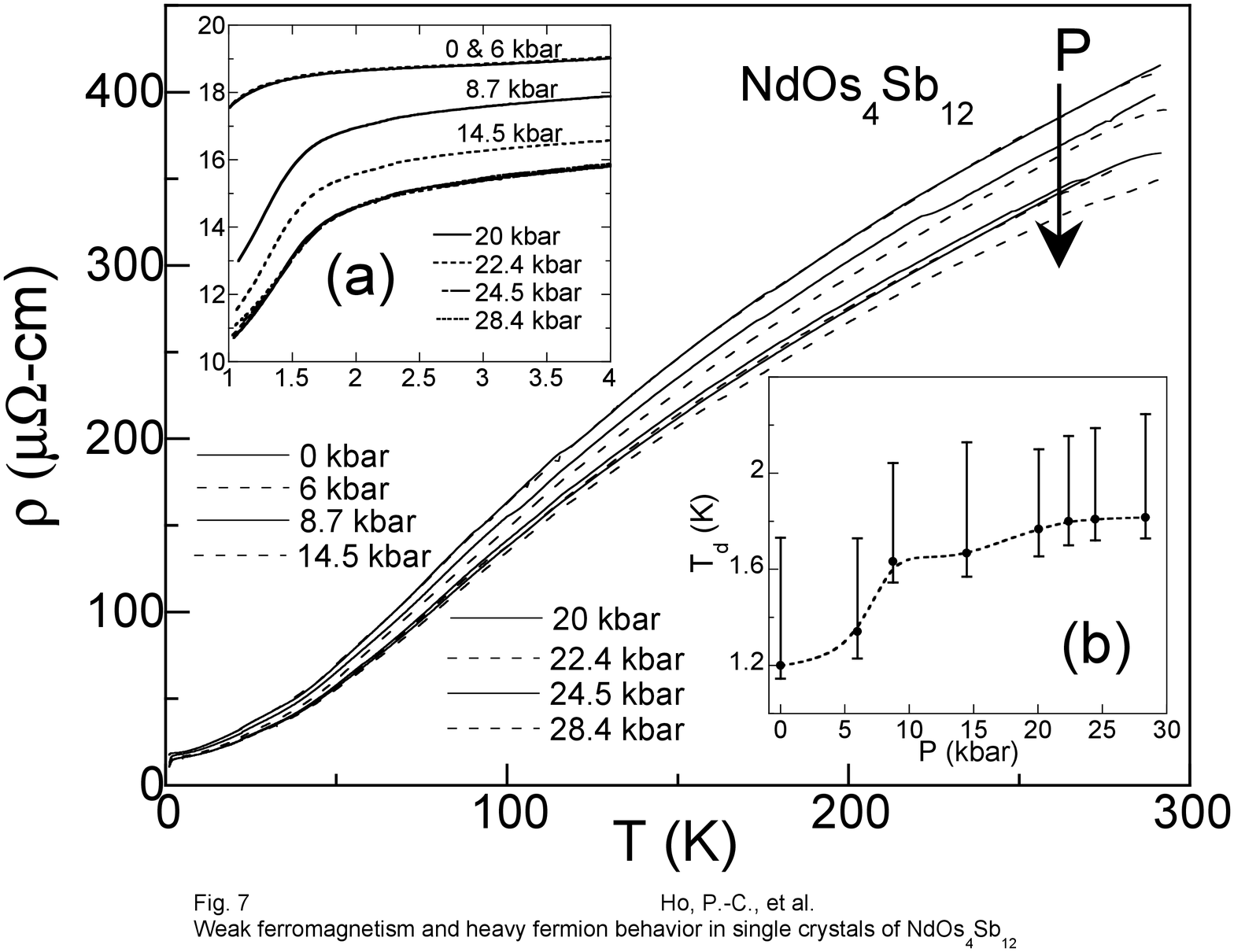}
 \caption{Electrical resistivity $\rho$ vs temperature $T$ for NdOs$_4$Sb$_{12}$
          at various pressures $P$ up to 28\,kbar.
          Inset (a): Expanded view of the resistive ferromagnetic transitions.
          Inset (b): Temperature of the drop in $\rho(T)$ due to the onset of ferromagnetism
                     $T_{\rm{d}}$ (approximately corresponding to  the Curie temperature
                     $\Theta_{\rm{C}}$) vs $P$ with vertical bars indicating the width
                     of the transition.}
 \label{fig:rhovsTinP}
\end{figure}

Measurements of $\rho(T)$ were performed from 1 K to 300 K under
nearly hydrostatic pressure $P$ between 0 kbar and 28 kbar
(Fig.~\ref{fig:rhovsTinP}).  In Fig.~\ref{fig:rhovsTinP} it can
clearly be seen that the pressure-induced change in the high-$T$
electrical resistivity is much more pronounced than the
resistivity change induced by high field
(Fig.~\ref{fig:rhovsTinH}(c)), in the pressure and temperature
ranges of this investigation. However, at low temperatures, the
value of $T_{\rm{d}}$ is more strongly influenced by an increase
in field $H$ (Fig.~\ref{fig:rhovsTinH}(b)) than by variation of
$P$ (Fig.~\ref{fig:rhovsTinP} inset (b)).

\subsection{Specific Heat}
\begin{figure}
 \epsfbox{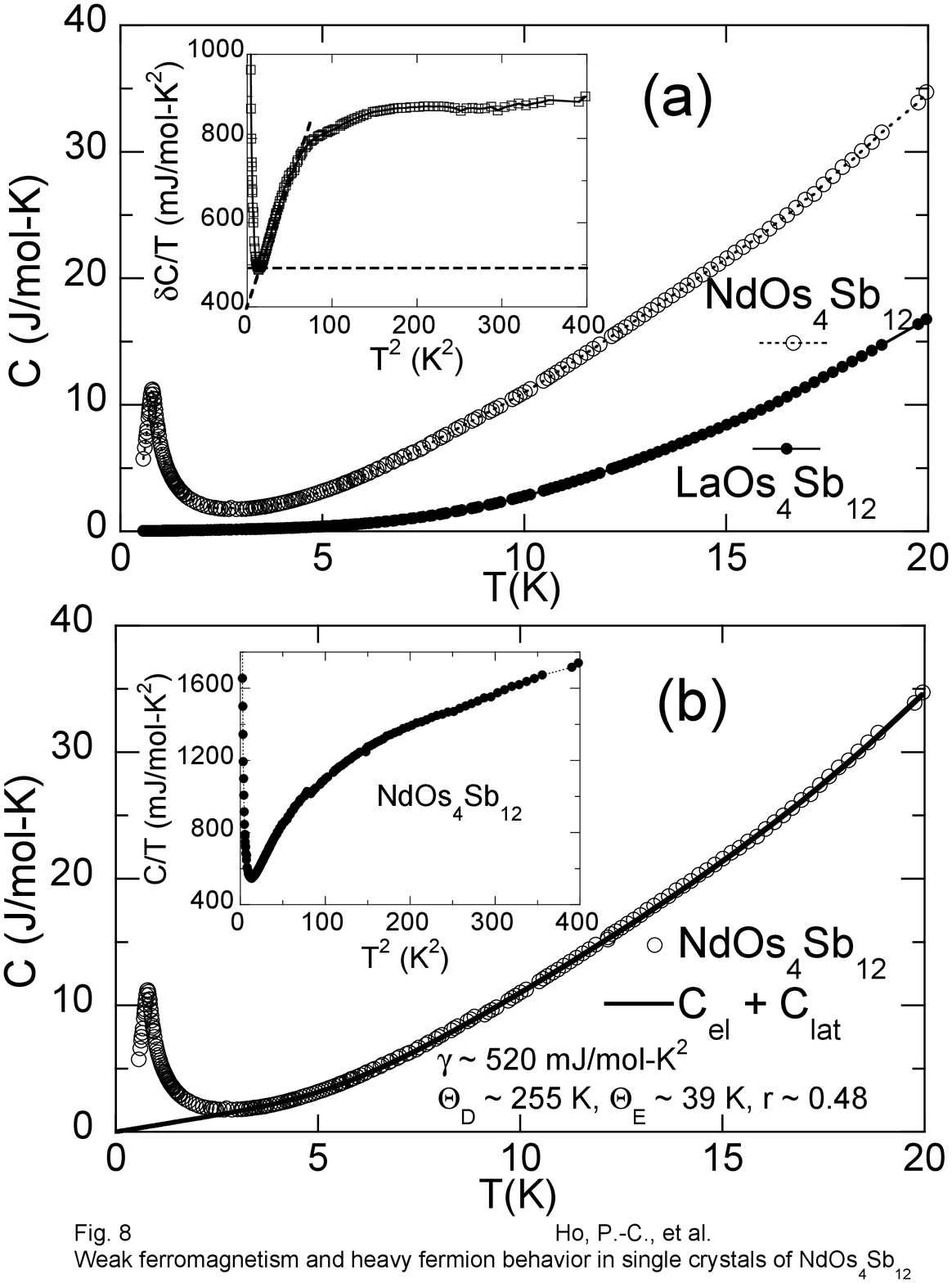}
 \caption{(a) Zero-field $C$ vs $T$ for NdOs$_4$Sb$_{12}$ and LaOs$_4$Sb$_{12}$ below 20\,K.
              Inset: $\delta C/T$ vs $T^2$ below 20\,K, where
              $\delta C \equiv C(\rm{NdOs_4Sb_{12}})$ -
              $C(\rm{LaOs_4Sb_{12}})$. The extrapolated values of the two dashed lines
              at 0\,K set the lower and upper limits of \mbox{$\gamma$(NdOs$_4$Sb$_{12}$) -
              $\gamma$(LaOs$_4$Sb$_{12}$)}.
          (b) A comparison between $C$ of NdOs$_4$Sb$_{12}$ and a fit of
              $C_{\rm{el}}$ + $C_{\rm{lat}}$, where $C_{\rm{el}} = \gamma T$
              and $C_{\rm{lat}}$ is composed of $C_{\rm{Deb}}$ and $C_{\rm{Ein}}$ as described in
              the text. The electronic specific heat coefficient $\gamma$,
              the Debye temperature ($\Theta_{\rm{D}}$), the Einstein temperature
              ($\Theta_{\rm{E}}$),and the coupling constant $r$ for the Einstein oscillator are estimated as
              520\,mJ/mol-K$^2$, 255\,K, 39\,K and 0.48 respectively.
              Inset: $C/T$ vs $T^2$ for NdOs$_4$Sb$_{12}$ below 20\,K for
              NdOs$_4$Sb$_{12}$.
          }
 \label{fig:CvsT@0T}
\end{figure}

Specific heat divided by temperature $C/T$ vs $T$ data for
NdOs$_4$Sb$_{12}$ are shown in Fig.~\ref{fig:CvsT@0T}(a). The data
reveal a pronounced peak at $\sim$ 0.8\,K that correlates well
with the magnetic ordering temperature $\Theta_{\rm{C}}$ inferred
from the shoulder of $\rho(T)$, the divergence of
$\chi_{\rm{dc}}(T)$, and the Arrott-plot analysis. No obvious
Schottky anomaly associated with CEF energy splitting was observed
in the $C(T)$ data below 20\,K.  This is consistent with the fact
that all the fits to the dc magnetic susceptibility data imply
that the CEF splitting between the ground and first excited states
is greater than 120\,K. The Schottky anomaly due to 120\,K CEF
splitting would exhibit a peak at \mbox{$\sim$ 40\,K} and make a
negligible contribution to $C(T)$ 20\,K.  Such a peak would also
be difficult to resolve against the large lattice background.

\epsfxsize=220pt \epsfysize=270pt

In Fig.~\ref{fig:CvsT@0T}(a), the specific heat of
LaOs$_4$Sb$_{12}$ is displayed in comparison with that of
NdOs$_4$Sb$_{12}$, where the electronic specific heat coefficient
$\gamma$ and the Debye temperature $\Theta_{\rm{D}}$ of
LaOs$_4$Sb$_{12}$ are \mbox{$\sim$ 36\,mJ/mol-K$^2$} and
\mbox{$\sim$ 280\,K}, respectively. After the specific heat of
nonmagnetic LaOs$_4$Sb$_{12}$ (an estimate of the lattice heat
capacity of NdOs$_4$Sb$_{12}$) is subtracted from the specific
heat of NdOs$_4$Sb$_{12}$, the difference divided by temperature
$\delta C/T$ is plotted against $T^2$ and shown in the inset of
Fig.~\ref{fig:CvsT@0T}(a). The value of $\gamma$ for
NdOs$_4$Sb$_{12}$ estimated from the $\delta C/T$ vs $T^2$ plot
ranges from \mbox{$\sim$ 436\,mJ/mol-K$^2$} to \mbox{$\sim$
530\,mJ/mol-K$^2$}. Below 13\,K, $\delta C/T$ vs $T^2$ is not
constant, which could be due to a difference between the actual
lattice heat capacities of NdOs$_4$Sb$_{12}$ and
LaOs$_4$Sb$_{12}$. Nevertheless, the curvature in $C/T$ vs $T^2$
of NdOs$_4$Sb$_{12}$ (inset of Fig.~\ref{fig:CvsT@0T}(b)) and the
magnetic transition occurring at $\sim$ 1\,K cause difficulties in
the analysis of the data using the typical formula $C/T =\gamma +
\beta T^2$. Since we have strong evidence against the possibility
of a CEF Schottky contribution from the analysis of the $\chi(T)$
data, the curvature in $C/T$ vs $T^2$ is possibly due to either
the rattling motion of the Nd atoms or a narrow peak in the
density of electronic states near Fermi energy, such as a Kondo
resonance.~\cite{Schotte1975} However, we do not consider the
application of the resonance level model (RLM)~\cite{Schotte1975}
to be appropriate, because the magnetization data above
$\Theta_{\rm{C}}$ are fit well by a Brillouin function and no
obvious Kondo effect is observed in the $\rho(T)$ data of
NdOs$_4$Sb$_{12}$. In the previous specific heat studies of the
Tl$_{0.22}$Co$_4$Sb$_{12}$ filled skutterudite compound, Sales et
al.~\cite{Sales1999} found that the difference in heat capacity
between Tl$_{0.22}$Co$_4$Sb$_{12}$ and the unfilled skutterudite
compound Co$_4$Sb$_{12}$ can be accurately described by a
quantized oscillator (Einstein model) with an Einstein temperature
$\Theta_{\rm{E}}$ of 55\,K. Since the X-ray structural refinement
at room temperature for NdOs$_4$Sb$_{12}$ indicates a small
$\Theta_{\rm{E}} \sim $ 45\,K associated with the Nd atoms, it can
be assumed that the Nd atoms partially act like Einstein
oscillators with a mixing ratio $r$, and the lattice contribution
to the specific heat can be expressed as $C_{\rm{lat}} =
C_{\rm{Ein}} + C_{\rm{Deb}}$,
\begin{equation}
C_{\rm{Ein}} = r \left[3 \rm{R} \frac{(\Theta_{\rm{E}}/\it{T})^2
e^{(\Theta_{\rm{E}}/\it{T})}} {\lbrack
e^{(\Theta_{\rm{E}}/\it{T})}-1\rbrack^2} \right],
\end{equation}
\begin{equation}
C_{\rm{Deb}} = (17-r) \cdot \frac{12\pi^4}{5} \rm{R} \left(
\frac{\it{T}}{\Theta_D} \right) ^3,
\end{equation}
where R is the universal gas constant, $\Theta_{\rm{D}}$
represents the Debye temperature, and \mbox{$r \leq  1$}. The
effective Debye contribution is $(17-r)/r$ times bigger than that
of the Einstein-like Nd rattling motion due to the participation
of the rest of the atoms in the unit cell. Below 20\,K, a least
squares fit of $C_{\rm{el}} + C_{\rm{lat}}$ to the $C(T)$ data was
performed, where $C_{\rm{el}} = \gamma T$ is the electronic
specific heat, from which estimated values of $\gamma,
\Theta_{\rm{D}}, \Theta_{\rm{E}}$, and $r$ were derived. The
fitting curve is plotted in Fig.~\ref{fig:CvsT@0T}(b) along with
the $C(T)$ data of NdOs$_4$Sb$_{12}$ as a comparison. The values
obtained for $\gamma, \Theta_{\rm{D}}$, $\Theta_{\rm{E}}$, and $r$
are estimated as 520\,mJ/mol-K$^2$, 255\,K, 39\,K, and 0.48,
respectively. The value of $\Theta_{\rm{D}}$ is close to the Debye
temperature of LaOs$_4$Sb$_{12}$, the value of $\Theta_{\rm{E}}$
is comparable to that determined from the X-ray data, and the
value of $\gamma$ is close to that estimated from the simple
subtraction of the LaOs$_4$Sb$_{12}$ specific heat data,
suggesting that NdOs$_4$Sb$_{12}$ is possibly a heavy fermion
compound.

\epsfxsize=200pt \epsfysize=250pt
\begin{figure}
 \epsfbox{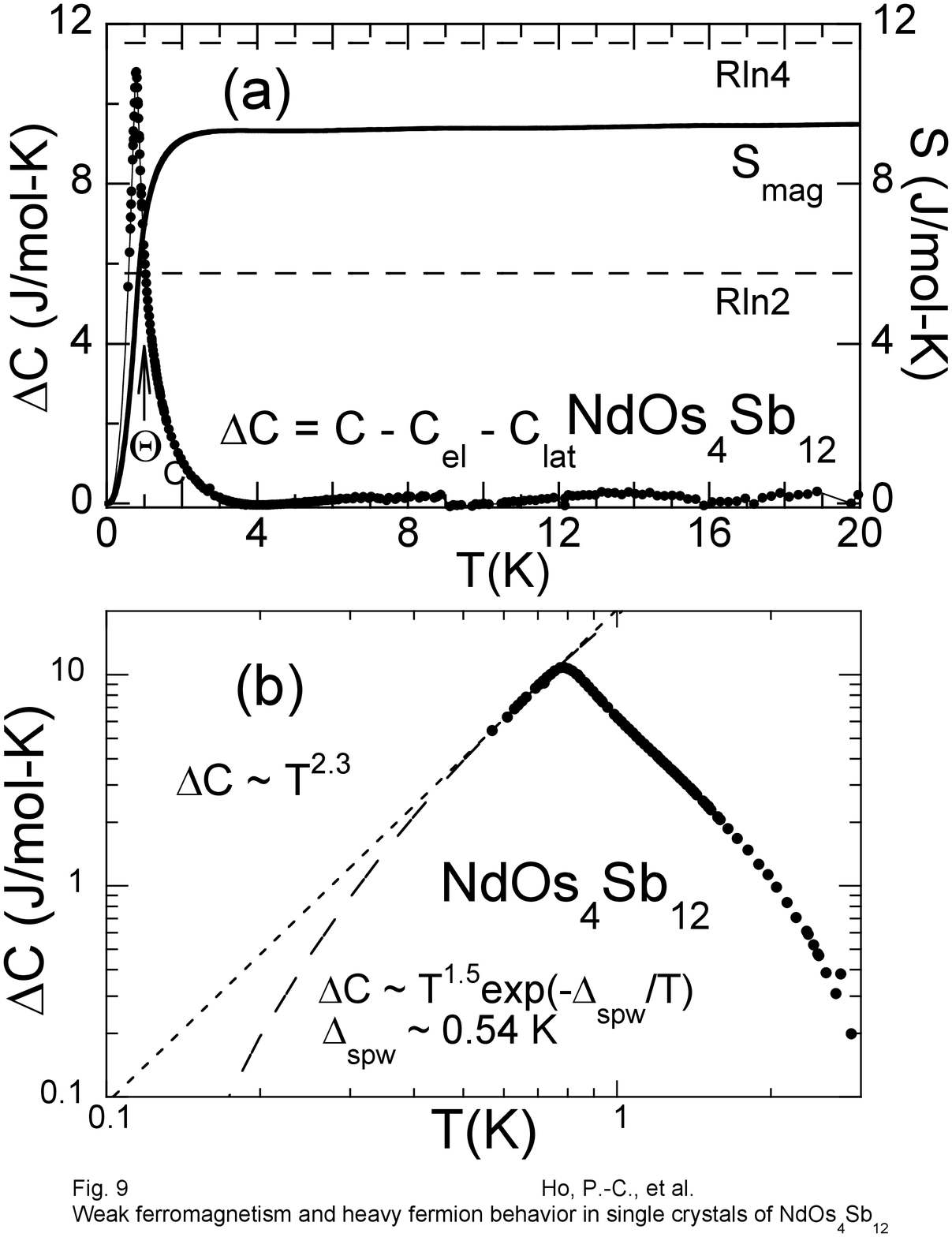}
 \caption{(a) Incremental specific heat $\Delta C$
          ($\Delta C \equiv C - C_{\rm{el}} - C_{\rm{lat}}$) (left axis)
          and the magnetic entropy $S_{\rm{mag}}$ (right axis) vs temperature $T$.
          (b) Logarithmic plot of the power-law fit (dotted line) and the
          anisotropic-spin-wave fit (dashed line) to the incremental specific heat $\Delta C(T)$
          after the electronic and lattice contributions have been removed (in a very
          limited measuring range below $\Theta_{\rm{C}}$).
         }
 \label{fig:DCandSvsT}
\end{figure}

The temperature dependence of the magnetic entropy $S_{\rm{mag}}$
was derived from the integration of $\Delta C/T$ vs $T$ and is
shown in Fig.~\ref{fig:DCandSvsT}(a), where $\Delta C \equiv C -
C_{\rm{el}} - C_{\rm{lat}}$. The magnetic entropy ($S_{\rm{mag}}$)
in NdOs$_4$Sb$_{12}$ reaches 0.69R ($\approx$ Rln2) at 0.85\,K and
levels off at a value of $\sim$ 1.14R ($\approx$ Rln3). The
magnetic entropy reaches \mbox{$\sim 74\%$} of its full value at
$\Theta_{\rm{C}}$, and a noticeable magnetic contribution persists
up to \mbox{$\sim$ 3\,K}, revealing the existence of magnetic
fluctuations above $\Theta_{\rm{C}}$. It has been argued for the
antiferromagnetic system Gd$_{1-x}$Y$_x$Ni$_2$Si$_2$ that magnetic
fluctuations can contribute to $C(T)$ at temperatures up to 5
times the N$\acute{\rm{e}}$el
temperature.~\cite{Sampathkumaran1995} Thus, we cannot completely
rule out the possibility that the short range magnetic
correlations near the ferromagnetic transition temperature regime
give rise to enhancement of the specific heat of
NdOs$_4$Sb$_{12}$.~\cite{Bouvier1991,Mallik1997,Sampathkumaran1995}
However, it is unlikely that they would account for a large
fraction of the enhanced specific heat, due to the following
arguments: (i) The paramagnetic-state $M(H)$ isotherm data (2 -
5\,K) scale well with a Brillouin function. The $M(H)$ data at
2\,K and 3\,K, and the $\chi_{\rm{dc}}^{-1}(T)$ data along with
the initial $\chi^{-1}$ from the Arrott Plot analysis do not show
obvious evidence of magnetic fluctuations above 2\,K (displayed in
Fig.~\ref{fig:ArrottPlot} (a) and (b)).  (ii) The value estimated
for $\gamma$ (520\,mJ/mol-K$^2$) is within the upper and lower
limits of $\gamma$ (436 - 530\,mJ/mol-K$^2$) estimated from the
analysis done by comparing the specific heat of NdOs$_4$Sb$_{12}$
with that of the nonmagnetic compound LaOs$_4$Sb$_{12}$ suggesting
that our analysis is justified. (iii) The lower limit (4\,K) of
the fitting range used to determine $\gamma$ from the formula $C =
C_{\rm{el}} + C_{\rm{lat}}$ is four times the Curie temperature,
which is safely higher than the temperature at which the
calculated $S_{\rm{mag}}$ saturates. (iv) Heavy fermion behavior
has been found in the neighboring compounds PrOs$_4$Sb$_{12}$
(\mbox{$\gamma \sim$ 600 mJ/mol-K$^2$})~\cite{Maple2003} and
SmOs$_4$Sb$_{12}$ (\mbox{$\gamma \approx$ 880
mJ/mol-K$^2$});~\cite{Yuhasz2004} earlier studies of the related
compound NdFe$_4$Sb$_{12}$ also reported possible evidence for an
enhanced electron mass.~\cite{Sugawara2000,Torikachvili1987}
Considering all of these points, it seems likely that
NdOs$_4$Sb$_{12}$ displays heavy fermion behavior.

Since $S_{\rm{mag}}$ lies between Rln2 (= 0.69R) and Rln4 (=
1.39R), it is difficult to determine conclusively whether the
$\Gamma_6$ doublet or $\Gamma_8^{(2)}$ quartet is the Nd$^{3+}$
ground state. If the $\Gamma_6$ doublet is the ground state, then
the extra entropy may result from another degree of freedom, such
as a tunnelling mode or off-center mode of Nd$^{3+}$ ion in an
Sb-icosahedron cage.~\cite{Goto2004} If the $\Gamma_8^{(2)}$
quartet is the ground state, the smaller entropy may be due to an
overestimated lattice contribution to the specific heat or
transfer of entropy to the conduction electron system.

Figure~\ref{fig:DCandSvsT}(b) displays the incremental specific
heat $\Delta C(T)$ after the electronic and lattice contributions
have been removed.  Even though the range of the $\Delta C(T)$
data below $\Theta_{\rm{C}}$ is very limited, the data were fit
with spin-wave formulas \mbox{$\Delta C(T) \propto T^{n}$} for
magnetically isotropic metals and \mbox{$\Delta C(T) \propto
T^{3/2}\exp(-\Delta_{\rm{spw}}/T)$} for magnetically anisotropic
metals.~\cite{Mackintosh1963} From the first formula, the value of
the exponent $n$ $\approx 2.3$ is higher than the value of 1.5
expected from a spin wave in an isotropic metal.  The spin-wave
energy gap $\Delta_{\rm{spw}}$ determined from the second formula
is $\sim 0.54$\,K, consistent with the value of 0.75\,K determined
from the zero-field resistivity data.

\section{Summary}
We have performed X-ray diffraction, electrical resistivity,
magnetization, and specific heat studies of NdOs$_4$Sb$_{12}$
single crystals, which exhibit interesting
strongly-correlated-electron behavior. X-ray experiments have
revealed full occupancy of Nd sites and a large mean square
displacement of Nd ions in NdOs$_4$Sb$_{12}$. The compound
NdOs$_4$Sb$_{12}$ exhibits mean-field-type ferromagnetism with
$\Theta_{\rm{C}} \sim$ 0.9\,K. The value of $\gamma$ estimated
from the analysis of the specific heat is large, $\sim
520$\,mJ/mol-K$^2$ (\mbox{$m^* \sim 98$\,$m_e$}), indicating that
NdOs$_4$Sb$_{12}$ is a possible heavy fermion compound. A cubic
CEF analysis suggests two best-fit energy level schemes: (I)
\mbox{$\Gamma_6$ (0\,K)}, \mbox{$\Gamma_8^{(1)}$ (180\,K)},
\mbox{$\Gamma_8^{(2)}$ (420\,K)}; (II) \mbox{$\Gamma_8^{(2)}$
(0\,K)}, \mbox{$\Gamma_8^{(1)}$ (220\,K)}, \mbox{$\Gamma_6$
(600\,K)}, both with a molecular field parameter \mbox{$\Lambda$ =
1.39 Nd-mol/cm$^3$}. The electrical resistivity data indicate that
both s-f exchange and aspherical Coulomb scattering are present in
NdOs$_4$Sb$_{12}$. Low-temperature electrical resistivity data
suggest the possible existence of spin-wave excitations below
$\Theta_{\rm{C}}$. The uncertainty in the CEF-energy-level scheme
ground state and the possible existence of spin-wave excitations
may be resolved by further neutron scattering experiments.

\begin{acknowledgments}
We thank Dr. C. Capan for technical support at NHMFL at Los Alamos
and Prof. D. P. Arovas at UCSD for helpful discussions. We also
thank S. K. Kim and A. P. Thrall for assistance in sample
preparation. Research at UCSD was supported by the \mbox{U. S.}
Department of Energy under Grant No. DE-FG02-04ER46105, the
\mbox{U. S.} National Science Foundation under Grant No.
DMR-0335173, and the National Nuclear Security Administration
under the Stewardship Science Academic Alliances program through
DOE Research Grant No. DE-FG52-03NA00068. The work at the NHMFL
Pulsed Field Facility (Los Alamos National Laboratory) was
performed under the auspices of the NSF, the State of Florida and
the US Department of Energy.
\end{acknowledgments}

\bibliographystyle{prsty}

\end{document}